\documentclass[iop,twocolappendix]{emulateapj}
\usepackage{amsmath}
\usepackage{apjfonts}

\newcommand{\be}{\begin{equation}}
\newcommand{\ee}{\end{equation}}
\newcommand{\bea}{\begin{eqnarray}}
\newcommand{\eea}{\end{eqnarray}}
\def\kms{\ {\rm km\, s}^{-1}}
\def\teff{T_{\rm eff}}
\def\logg{{\rm log}\,g}

\shortauthors{CONROY ET AL.}

\shorttitle{The H3 Survey}

\begin{document}


\title{Mapping the Stellar Halo with the H3 Spectroscopic Survey}

\author{Charlie Conroy\altaffilmark{1}, Ana Bonaca\altaffilmark{1},
  Phillip Cargile\altaffilmark{1}, Benjamin
  D. Johnson\altaffilmark{1}, Nelson Caldwell\altaffilmark{1}, Rohan
  P. Naidu\altaffilmark{1}, Dennis Zaritsky\altaffilmark{2}, Daniel
  Fabricant\altaffilmark{1}, Sean Moran\altaffilmark{1}, Jaehyon
  Rhee\altaffilmark{1}, Andrew Szentgyorgyi\altaffilmark{1}}

\altaffiltext{1}{Center for Astrophysics $\vert$ Harvard \& Smithsonian, Cambridge, MA, 02138, USA}
\altaffiltext{2}{Steward Observatory, University of Arizona, 933 North Cherry
  Avenue, Tucson, AZ 85721, USA}

\slugcomment{Submitted to ApJ}

\begin{abstract}

  Modern theories of galaxy formation predict that the Galactic
  stellar halo was hierarchically assembled from the accretion and
  disruption of smaller systems.  This hierarchical assembly is
  expected to produce a high degree of structure in the combined phase
  and chemistry space; this structure should provide a relatively
  direct probe of the accretion history of our Galaxy. Revealing this
  structure requires precise 3D positions (including distances), 3D
  velocities, and chemistry for large samples of stars.  The {\it
    Gaia} satellite is delivering proper motions and parallaxes for
  $>1$ billion stars to $G\approx20$.  However, radial velocities and
  metallicities will only be available to $G\approx15$, which is
  insufficient to probe the outer stellar halo ($\gtrsim10$ kpc).
  Moreover, parallaxes will not be precise enough to deliver
  high-quality distances for stars beyond $\sim10$ kpc.  Identifying
  accreted systems throughout the stellar halo therefore requires a
  large ground-based spectroscopic survey to complement {\it Gaia}.
  Here we provide an overview of the H3 Stellar Spectroscopic Survey,
  which will deliver precise stellar parameters and spectrophotometric
  distances for $\approx200,000$ stars to $r=18$.  Spectra are
  obtained with the Hectochelle instrument at the MMT, which is
  configured for the H3 Survey to deliver resolution $R\approx23,000$
  spectra covering the wavelength range $5150$\AA$-5300$\AA. The
  survey is optimized for stellar halo science and therefore focuses
  on high Galactic latitude fields ($|b|>30^\circ$), sparsely sampling
  $15,000$ sq. degrees.  Targets are selected on the basis of {\it
    Gaia} parallaxes, enabling very efficient selection of bone fide
  halo stars.  The survey began in the Fall of 2017 and has collected
  88,000 spectra to-date.  All of the data, including the derived
  stellar parameters, will eventually be made publicly available via
  the survey website: \texttt{h3survey.rc.fas.harvard.edu}.

\end{abstract}

\keywords{Galaxy: halo --- Galaxy: kinematics and dynamics}


\section{Introduction \& Motivation}
\label{s:intro}

Simulations of hierarchical structure formation predict that the
stellar halo contains an extraordinary amount of structure in the high
dimensional space (6D phase plus N dimensional abundances plus stellar
ages) occupied by stars \citep[e.g.,][]{Johnston96, Helmi99,
  Bullock05, Cooper10}.  Thankfully, historical memory in the Galactic
halo stretches back billions of years due to long relaxation times;
one of the principal goals of {\it Galactic Archeology} is to uncover
this history and use it to trace the assembly of our Galaxy
\citep[e.g.,][]{Helmi08}.  Learning this assembly history impacts many
fields of astronomy, including galaxy formation, the cosmological
evolution of structure, and studies of the nature of dark matter.

\begin{figure*}[!t]
\center
\includegraphics[width=0.95\textwidth]{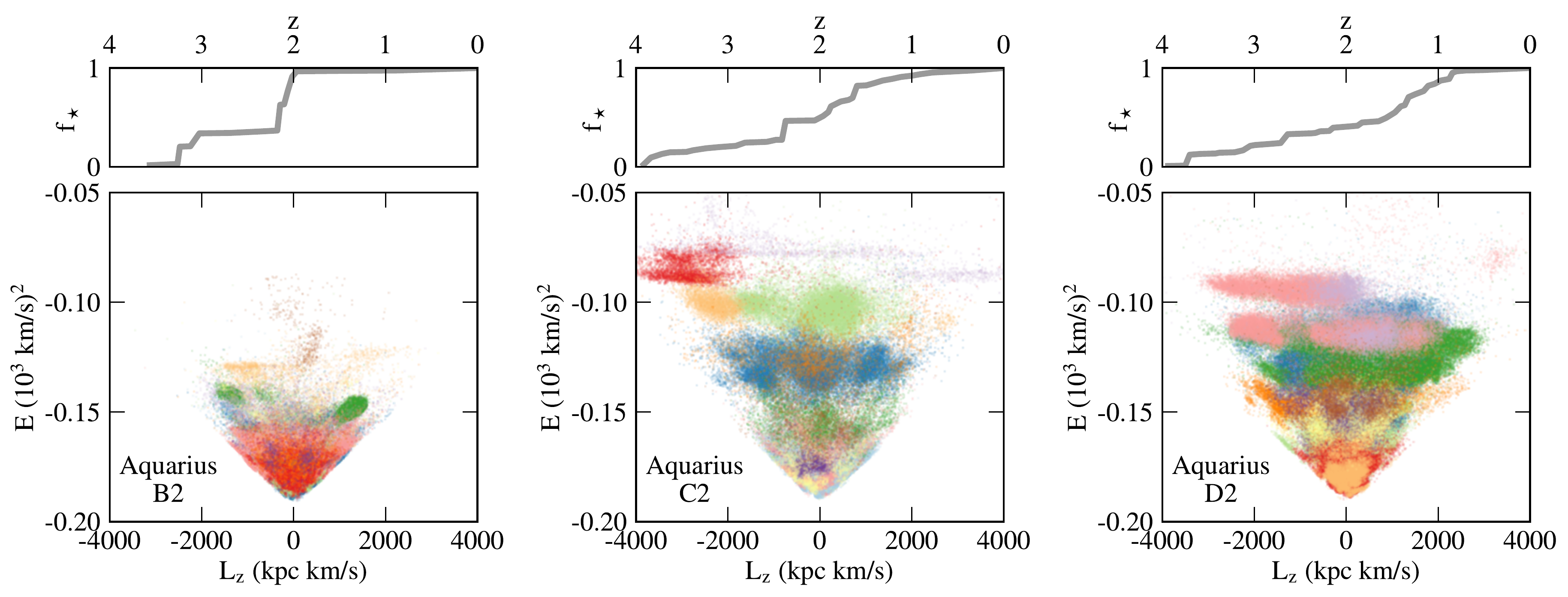}
\vspace{0.1cm}
\caption{Phase space encodes the assembly history of the stellar halo.
  The three panels show the distribution in energy-angular momentum
  ($E-L_Z$) space of stellar halos from the Aquarius simulations
  (bottom panels), along with the stellar halo assembly histories (top
  panels).  Unique progenitors are assigned distinct colors in the
  bottom panels.  Notice that relatively early assembly histories
  (left panel) results in a notably different phase space distribution
  compared to late assembly (right panel).}
\label{fig:elzcomp}
\end{figure*}

The distribution of stars in energy and angular momentum, e.g.,
$E-L_Z$ space, encodes the accretion history of galaxies, as
demonstrated in Figure \ref{fig:elzcomp}.  In this figure we show
three simulated stellar halos with both ``early'' and ``late''
accretion histories \citep[from the Aquarius simulations;][]{Cooper10,
  Lowing15}.  The distribution of stars in these model halos in energy
and angular momentum varies systematically with accretion history.
The measurement of the phase space structure of the Galactic stellar
halo should therefore place novel constraints on the Galaxy's assembly
history.

The {\it Gaia} mission \citep{Perryman01} is propelling a revolution
in our understanding of the stellar halo \citep[e.g.,][]{Helmi18,
  Belokurov18}.  {\it Gaia} is delivering parallaxes and proper
motions for $>1$ billion stars to $G\approx20$.  In DR2
\citep{GaiaDR2}, radial velocities (RVs) were measured for stars
brighter than $G\approx12$; the end-of-mission projection predicts
that RVs with precision $5-10\kms$ will be available to $G\approx15$.
This leaves a significant gap, of $\approx5$ mag, where {\it Gaia}
will provide constraints on parallaxes and proper motions for stars
lacking RVs.  This gap must be filled by large ground-based
spectroscopic surveys.

The need for spectroscopic information is motivated in Figure
\ref{fig:motiv}.  Here we show a simulated stellar halo from the
Aquarius simulation halo A2.  In the upper panels we show all of the
stars in grey and highlight in orange the stars belonging to a single
progenitor.  The upper left panel shows a sky projection in the
Northern hemisphere.  It is clear that debris from a single progenitor
is spread across the sky, necessitating large area surveys.  The right
panel shows a projection of phase space: energy and the $z-$component
of the angular momentum, $E-L_Z$, in which many structures are clearly
visible.  The middle panel shows pseudo $E-L_Z$ space, in which no
radial velocities are available and so we have assumed RV$=0.0\kms$.
While some of the largest structures are still visible, most of the
structure is washed away.  The bottom panel shows $E-L_Z$ space
color-coded by stellar metallicity.

Clearly, a spectroscopic survey that can deliver RVs and metallicities
for stars fainter than $G\approx15$, would be highly complementary to
the {\it Gaia} spectroscopic data.  The faintest dwarfs have internal
velocity dispersions of order a few $\kms$ \citep{Simon07}, so one
would need RVs with a precision of $\lesssim 1 \kms$.  The bottom
panel of Figure \ref{fig:motiv} suggests a measurement precision on
[Fe/H] of $\sigma_{\rm [Fe/H]}\sim 0.05$ dex should be sufficient to
aid in the identification of structure in phase space.

For stars fainter than $G\approx15$ beyond $\sim10$ kpc, {\it Gaia}
parallaxes are too noisy to determine precision distances.  Therefore,
spectrophotometric distances \citep[e.g.,][]{Burnett10} are also
critically important in order to study the stellar halo.  Figure
\ref{fig:elzdist} shows the Aquarius A2 halo from Figure
\ref{fig:motiv} as a function of the distance precision.  Large
uncertainties can have a dramatic effect on the structure in phase
space, and uncertainties typically scatter stars along diagonal
directions in $E-L_Z$ space.

\begin{figure}[t!]
\center
\includegraphics[width=0.48\textwidth]{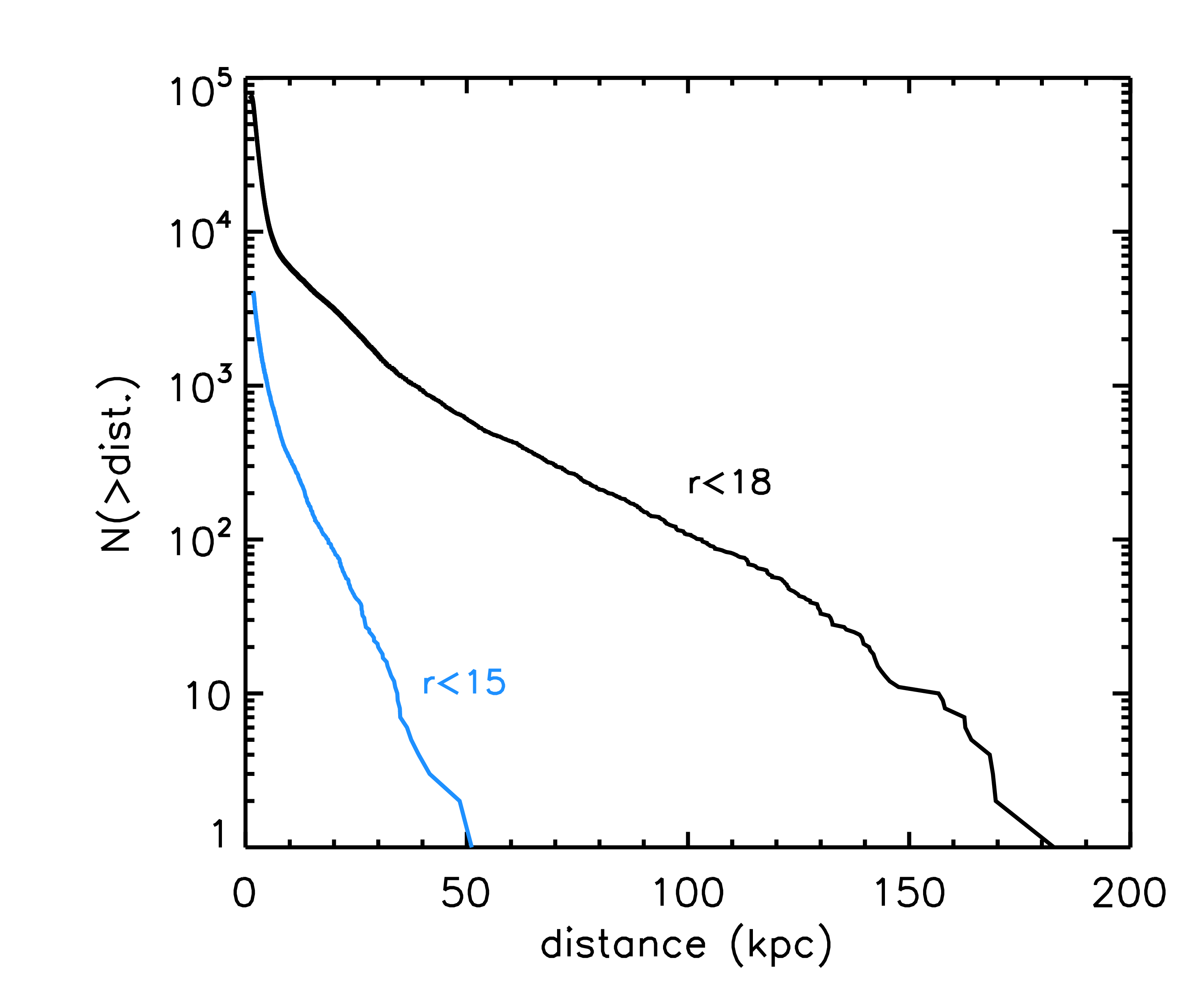}
\vspace{0.1cm}
\caption{Cumulative distribution of stellar distances in the
  \citet{Rybizki18} mock galaxy catalog for stars brighter than a
  magnitude limit of $r=15$ and $r=18$.  Only stars located within the
  existing H3 Survey footprint are included (i.e., $|b|>40^\circ$ and
  Dec.$>-20^\circ$).  A limiting magnitude of $\approx15$ is
  insufficient to sample the outer stellar halo, as even stars at the
  tip of the red giant branch are too dim to be seen beyond
  $\approx50$ kpc.  Note that the {\it Gaia} $G-$band is similar to
  $r$ (to within $\approx0.05$ mag) for 4000 K $<\teff<6500$ K.}
\label{fig:maglim}
\end{figure}

Finally, Figure \ref{fig:maglim} shows the cumulative distribution of
stellar distances in the \citet{Rybizki18} mock Milky Way catalog for
a limiting magnitude of $r=18$ and $r=15$.  This mock catalog assumes
smooth models for the stellar disk and halo.  Here we plot stars
selected to have a parallax $\pi<0.5$ mas, and to lie within the
R.A.-Dec. footprint shown in Figure \ref{fig:footprint} (these details
do not affect the main point of Figure \ref{fig:maglim}).  It is
clear that a relatively bright limit of $\approx15$ mag is
insufficient to sample the outer stellar halo.

In summary, a ground-based spectroscopic complement to {\it Gaia} that
aims to study the entire stellar halo should meet the following
requirements: 1) accurate RVs ($\sigma_{\rm RV}\lesssim 1\kms$),
spectrophotometric distances (to $\lesssim10$\%), and metallicities
(at a precision of $\lesssim 0.05$ dex); 2) a depth of $\sim18$ mag;
3) cover a large fraction of the sky.  These considerations are the
motivation behind the H3 Survey (``Hectochelle in the Halo at High
Resolution''), which we describe in detail below.  The survey is
driven by the following science cases:

\begin{figure*}[!t]
\center
\includegraphics[width=0.95\textwidth]{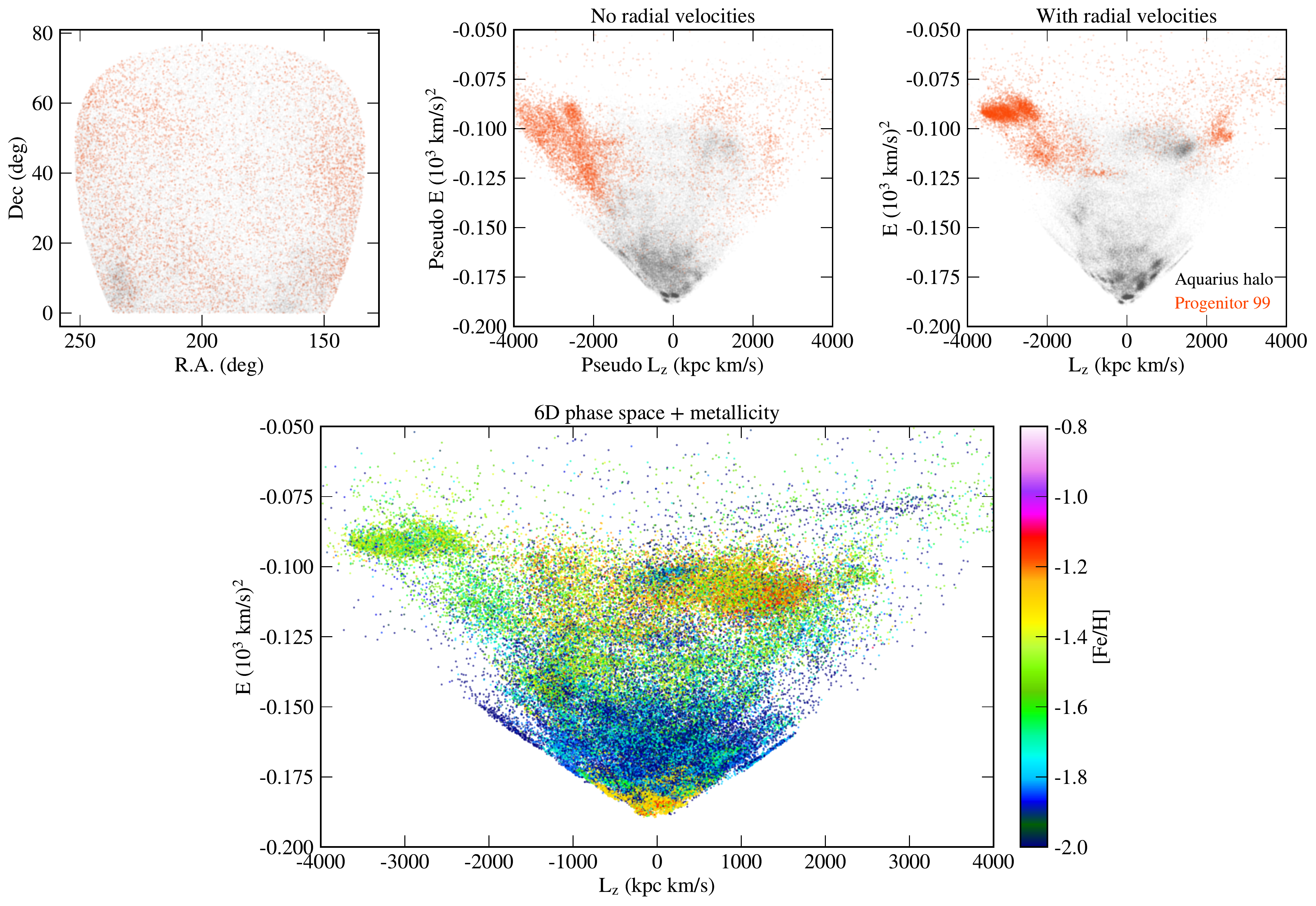}
\vspace{0.1cm}
\caption{Importance of radial velocities (RVs) and metallicities for
  revealing structure in phase space.  A simulated stellar halo
  (Aquarius A2) is shown in sky coordinates (upper left), $E-L_Z$
  space both with (upper right) and without (upper middle) RVs.  In
  the latter case the quantities are estimated assuming $RV=0.0$.  The
  lower panel shows the additional information provided by stellar
  metallicities.  The top set of panels highlights in orange a single
  massive progenitor.  The debris of such a system is spread
  throughout the sky, necessitating spectroscopic observations over a
  large fraction of the sky.  Notice the addition of radial velocities
  (upper middle vs. upper right) results in much sharper structures in
  $E-L_Z$ space.}
\label{fig:motiv}
\end{figure*}

\begin{enumerate}

\item {\it Identify and characterize disrupting and disrupted dwarf
    galaxies throughout the Milky Way stellar halo using both phase
    space information and chemistry.}  As shown in Figures
  \ref{fig:elzcomp} and \ref{fig:motiv}, phase space is rich with
  substructure that reflects the accretion and disruption of many
  dwarf galaxies over the history of the Galaxy.  Recent work using
  {\it Gaia} DR2 data has identified one, or possibly more, accreted
  galaxies in the inner halo \citep[e.g.,][]{Helmi18, Myeong18a,
    Fattahi19}, although the interpretation of discrete structures in
  phase space is challenging \citep{Jean-Baptiste17}.

\item {\it Fully characterize the stellar halo in order to understand
    the accretion history of the Galaxy.}  Figure \ref{fig:elzcomp}
  demonstrates that ``early'' vs. ``late'' accretion histories result
  in qualitatively different distributions of stars in phase space.
  There are few if any constraints available regarding the accretion
  history of our Galaxy and the phase space structure offers one of
  the strongest discriminants \citep[e.g.,][]{Deason13}.

\item {\it Stellar populations in the halo.}  There are a wide variety
  of interesting and unusual stellar populations found in the halo
  including carbon-enhanced metal poor (CEMP) stars \citep{Lee13},
  high velocity stars, and very metal-poor stars \citep{Frebel15}.
  These populations offer valuable clues to the origin of the stellar
  halo and the structure of the Galaxy.

\item {\it Establish the origin of the in-situ stellar halo}.  Recent
  hydrodynamic simulations and analysis of {\it Gaia} DR1 data suggest
  that there may be an {\it in-situ} stellar halo (an ``inner'' halo)
  formed from scattered disk stars \citep[e.g.,][]{Bonaca17}.
  Determining the formation mechanism of an {\it in-situ} stellar halo
  will provide unique clues to the early Galaxy and will place
  constraints on major disruptive events in its dynamical history.

\item {\it Measure the outer mass profile of the Galaxy and the shape
    of the dark matter halo.}  Analysis of stellar streams has
  provided novel but often conflicting conclusions regarding the shape
  and mass profile of the Galaxy.  The total mass of the Galaxy is
  unknown to a factor of two \citep{Wang15b}, which is a major limiting
  factor when placing the Galaxy in a cosmological context.  Various
  outstanding puzzles, including the ``missing satellites'' and ``too
  big to fail'' problems become more or less severe depending on the
  mass of the Galaxy's halo \citep[e.g.,][]{Bullock17}.  Velocities
  and distances in the outer stellar halo will enable strong
  constraints on the mass of the Galaxy's dark halo.

\item {\it Measure the mass of the Large Magellanic Cloud (LMC)
    through its gravitational effect on the stellar halo.}  Recent
  work has demonstrated that the impact of the LMC on the Galaxy can
  be substantial, both by inducing movement of the Galactic barycenter
  due to gravitational acceleration \citep{Gomez15,
    Garavito-Camargo19} and a more complicated change in stellar (and
  dark matter) densities, mean velocities, and velocity dispersions
  induced in the ``wake'' of the LMC's orbit
  \citep{Garavito-Camargo19}.  \citet{Belokurov19a} report a tentative
  detection of this wake in the direction of the Pisces Overdensity.
  The reflex motion of the Galactic barycenter can induce a dipole
  pattern in the RVs across the outer halo as large as $\pm50\kms$
  depending on the mass of the LMC.

\end{enumerate}

This paper is organized as follows.  In Section \ref{s:design} we
describe the design of the survey, including the footprint and target
selection.  Section \ref{s:obs} outlines the observing strategy, data
reduction, and stellar parameter pipeline.  Survey progress to-date is
presented in Section \ref{s:prog} and we place the survey in the
context of existing and future spectroscopic surveys in Section
\ref{s:context}.  A summary is provided in Section \ref{s:sum}.

\begin{figure*}[!t]
\center
\includegraphics[width=0.95\textwidth]{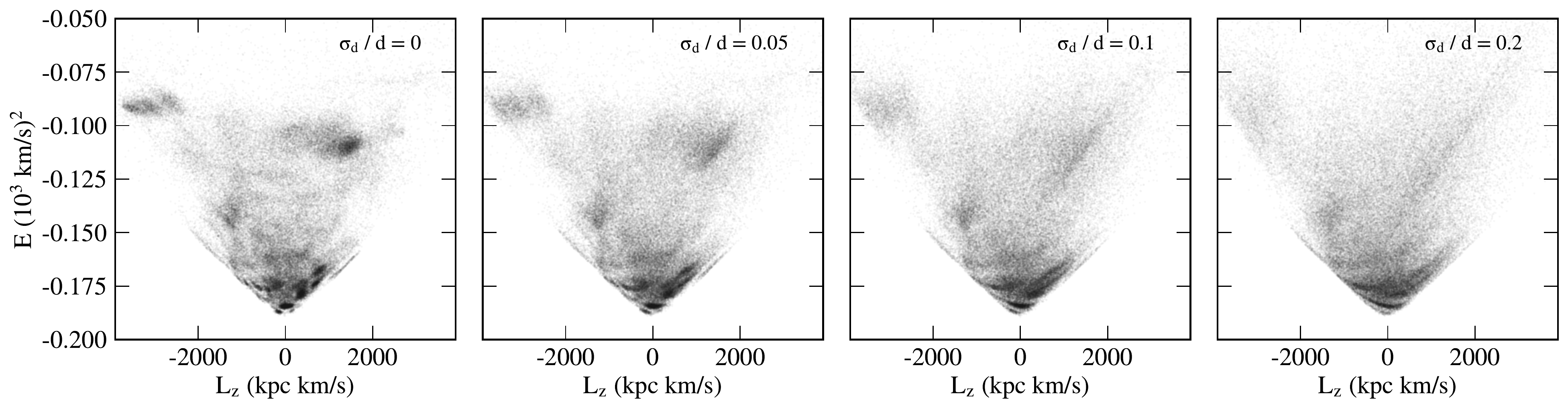}
\vspace{0.1cm}
\caption{Effect of distance uncertainties on the $E-L_Z$ diagram.  The
  simulated halo is the same as in Figure \ref{fig:motiv} (Aquarius
  A2).  The four panels show the impact of 0\%, 5\%, 10\%, and 20\%
  distance uncertainties.  Notice that many clumps in the leftmost
  panel turn into diagonal streaks in subsequent panels.}
\label{fig:elzdist}
\end{figure*}


\section{Survey Design}
\label{s:design}

\subsection{Survey fields}

As Figure \ref{fig:motiv} shows, the debris from accreted galaxies is
typically spread over large fractions of the sky.  A survey aimed at
identifying and studying the various components of the stellar halo
must therefore cover a large area.  Guided by this consideration, our
survey strategy is to sparsely sample the entire sky observable from
the MMT, while avoiding the Galactic plane.  The latter constraint
ensures that the fields are not dominated by disk stars.  Our survey
footprint is therefore the area on the sky encompassing
Dec.$>-20^{\circ}$ and $|b|>30^{\circ}$, which totals $\approx15,000$
sq. degrees.  Within this footprint we define fields spaced every
$3^{\circ}$ in both R.A. and Dec.  There are 1654 such fields, of
which we intend to observe approximately 1,000.  We do not
specifically target, nor avoid, fields with known structures.

\begin{figure}[t!]
\center
\includegraphics[width=0.48\textwidth]{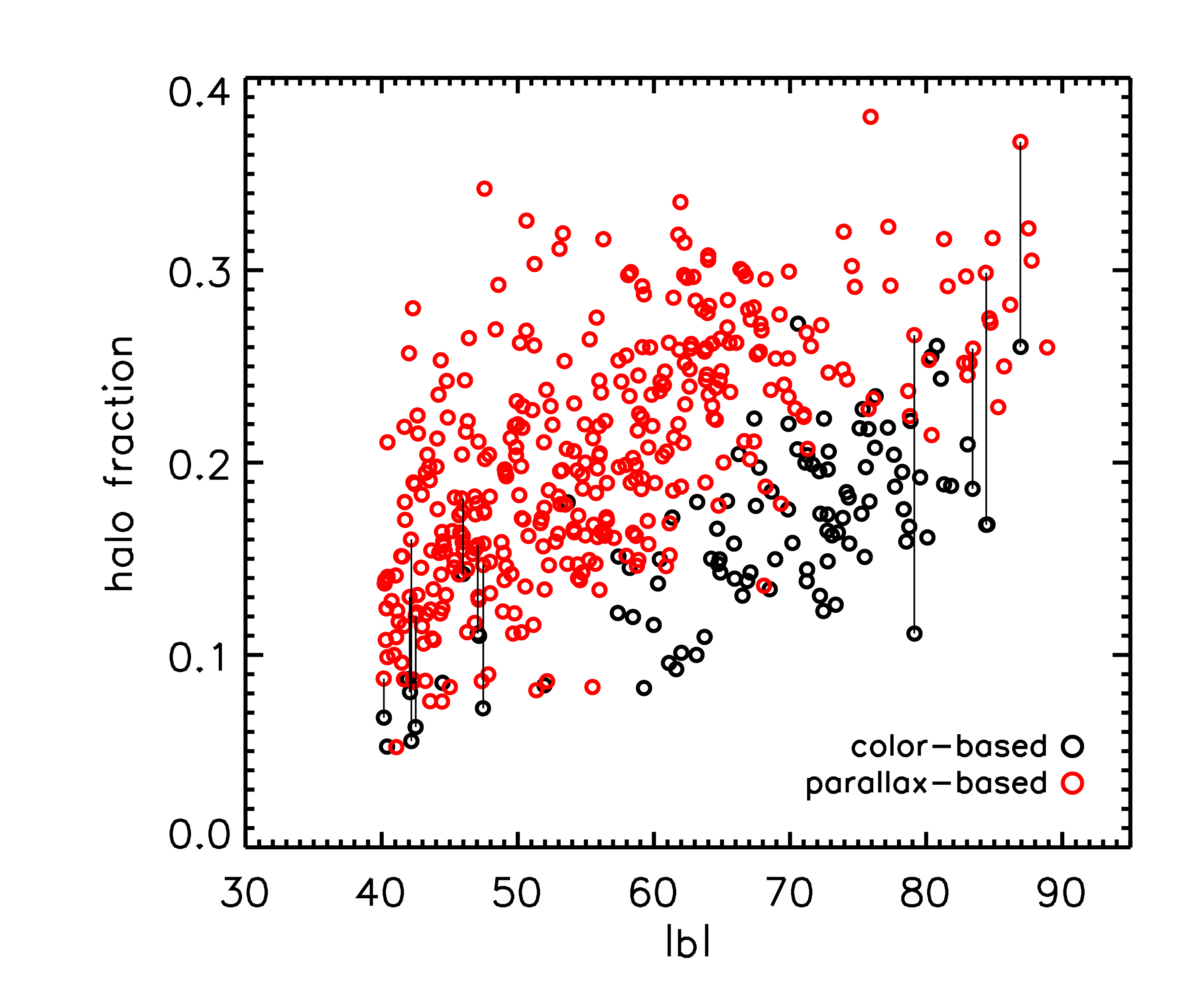}
\vspace{0.1cm}
\caption{Fraction of stars per field that belong to the kinematic
  stellar halo, as a function of Galactic latitude.  Black symbols
  show the early survey fields where the target selection was based on
  color-cuts.  Red symbols show fields where {\it Gaia} parallaxes
  were used to define the sample.  Lines connect the same fields
  observed with both selections.  Notice that our simple {\it Gaia}
  parallax selection results in a much higher fraction of halo stars
  compared to color-based selection.}
\label{fig:hfrac}
\end{figure}


\subsection{Input catalog \& target selection}
\label{s:sel}

The survey began collecting data in the Fall of 2017, before {\it
  Gaia} DR2 was available, so our input catalog is based on Pan-STARRS
data release 1 \citep[PS1;][]{Chambers16}.  This primary catalog was
matched to {\it WISE} \citep{Wright10, Cutri13}, SDSS, 2MASS, and,
later, {\it Gaia} DR2.  The cross-matching was performed using the
Large Survey Database framework \citep{Juric12} with a matching radius
of $<1\arcsec$.

AGN are identified and removed from the catalog with a {\it WISE}
color-cut: $0.85<W1-W2<2.0$.  We require stars to satisfy two {\it
  Gaia} quality cuts: \texttt{astrometric\_excess\_noise} $<1$ and
\texttt{visibility\_periods\_used} $>5$ \citep{Lindegren18}.

The over-arching goal of the target selection is to be as simple as
possible while optimizing for distant stars.  Prior to {\it Gaia} DR2
the main sample was defined by the following two requirements:
$15<r<18$ and $g-r<1$.  The latter is included to filter out the
numerous foreground cool dwarfs.  A total of 19,301 stars were
observed with this selection.  After the release of {\it Gaia} DR2 the
selection switched to a simple parallax cut, with the same magnitude
range of $15<r<18$.  Specifically, we first employed a parallax
selection to ensure that we included all stars that had parallaxes
within 2-sigma of being less than $0.5$ mas, i.e.,
$\pi-2\sigma_\pi<0.5$.  We recently modified this selection slightly
to deliver an even higher halo fraction by adopting $\pi<0.4$ mas as
the main sample selection.

There are known systematic issues with the {\it Gaia} DR2 parallaxes
at the $\lesssim0.1$ mas level.  \citet{Lindegren18} reported
zeropoint offsets based on quasars at the 0.03 mas level.  However,
they recommended against applying an overall zeropoint correction to
the reported parallaxes, noting substantial small and large-scale
spatial variation in the zeropoints.  Subsequent work
\citep[e.g.,][]{Stassun18, Schonrich19, Leung19} has confirmed the
existence of a global zeropoint offset, and has demonstrated that the
offset depends on magnitude and color.  Owing to the complex nature of
the corrections, we have not applied any correction to the DR2
parallaxes at the target selection stage (corrections are applied
later when estimating stellar parameters).  Hopefully this issue will
be resolved in DR3, at which time we will be able to quantify the DR2
zeropoint effect on the selection function with mock catalogs.

In addition to the main sample, we also include rare, high-value
targets.  We identify K giants from the input photometric catalogs via
a series of optical-IR color cuts described in \citet{Conroy18a}.
These stars are rare but the purity of the selection is very high
($\approx85$\% probability of being a giant; see Section \ref{s:prog})
and they are visible to $>100$ kpc at $r=18$.  We also select blue
horizontal branch (BHB) stars according to the color cuts presented in
\citet{Deason14}, and RR Lyrae from the catalog of \citet{Sesar17}.
These three categories of targets are rare, comprising only a few
stars per field.  They are given higher priority than the main sample
in assigning fibers to targets.

Two sets of filler targets are included in cases where there are not
enough targets in the main and high-value samples to fill the
available fibers.  The first set of fillers have the same selection as
the main sample but are fainter: $18<r<18.5$.  The second (lowest
priority) fillers have $15<r<18$ and $0.4<\pi<1.0$ mas.

Figure \ref{fig:hfrac} shows the resulting fraction of stars per field
that are identified as belonging to the kinematic stellar halo
(specifically, $|V-V_{\rm LSR}|>220 \kms$).  We compare the halo
fraction for the early color-based selection to the later
parallax-based color selection.  As noted below, 12 fields were
observed twice, once for each of the two selections.  Solid lines in
the figure connect the halo fractions for the same fields observed
with the two selection functions.  In all cases the halo fraction is
substantially higher with the parallax-based selection.  The median
halo fraction of the parallax-based selection is 20\%, with a strong
dependence on Galactic latitude (no fields at $30^\circ<|b|<40^\circ$
have been observed to-date).

\begin{figure*}[!t]
\center
\includegraphics[width=0.99\textwidth]{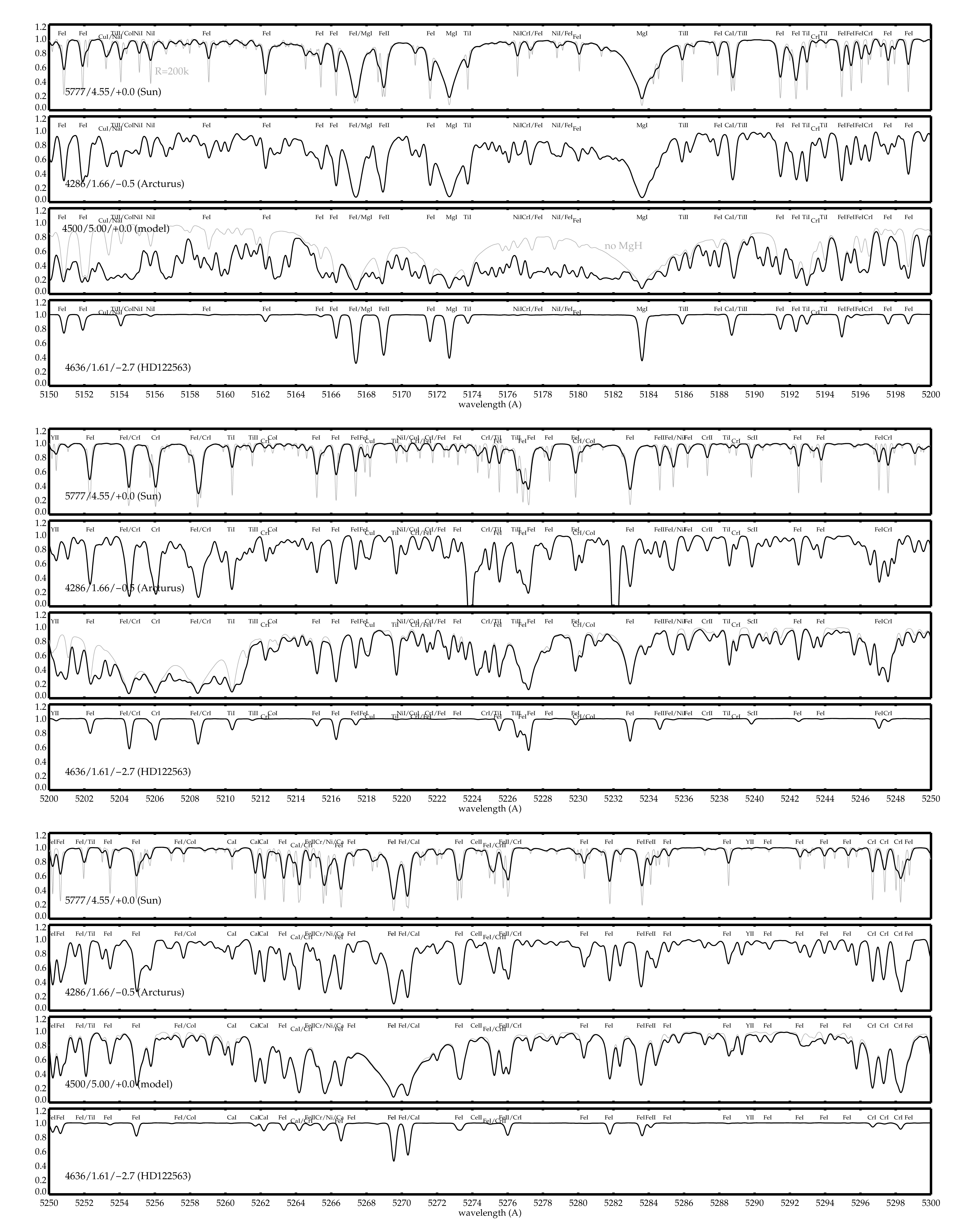}
\vspace{0.1cm}
\caption{Spectra of four example stars in the spectral region covered
  by the survey.  From top to bottom: spectra of the Sun, Arcturus, a
  model M dwarf, and a metal-poor giant (HD122563).  In each panel the
  stellar parameters are listed as $\teff$/$\logg$/[Fe/H].  Many of
  the strongest features are labeled.  All spectra are convolved to
  $R=23,000$.  In the top panel the $R\approx400,000$ solar spectrum
  is included as a grey line.  For the model M dwarf, we show in grey
  a spectrum computed without the MgH molecular features.}
\label{fig:rv31}
\end{figure*}


\section{Observations \& Data Analysis}
\label{s:obs}

\begin{figure*}[!t]
\center
\includegraphics[width=1.\textwidth]{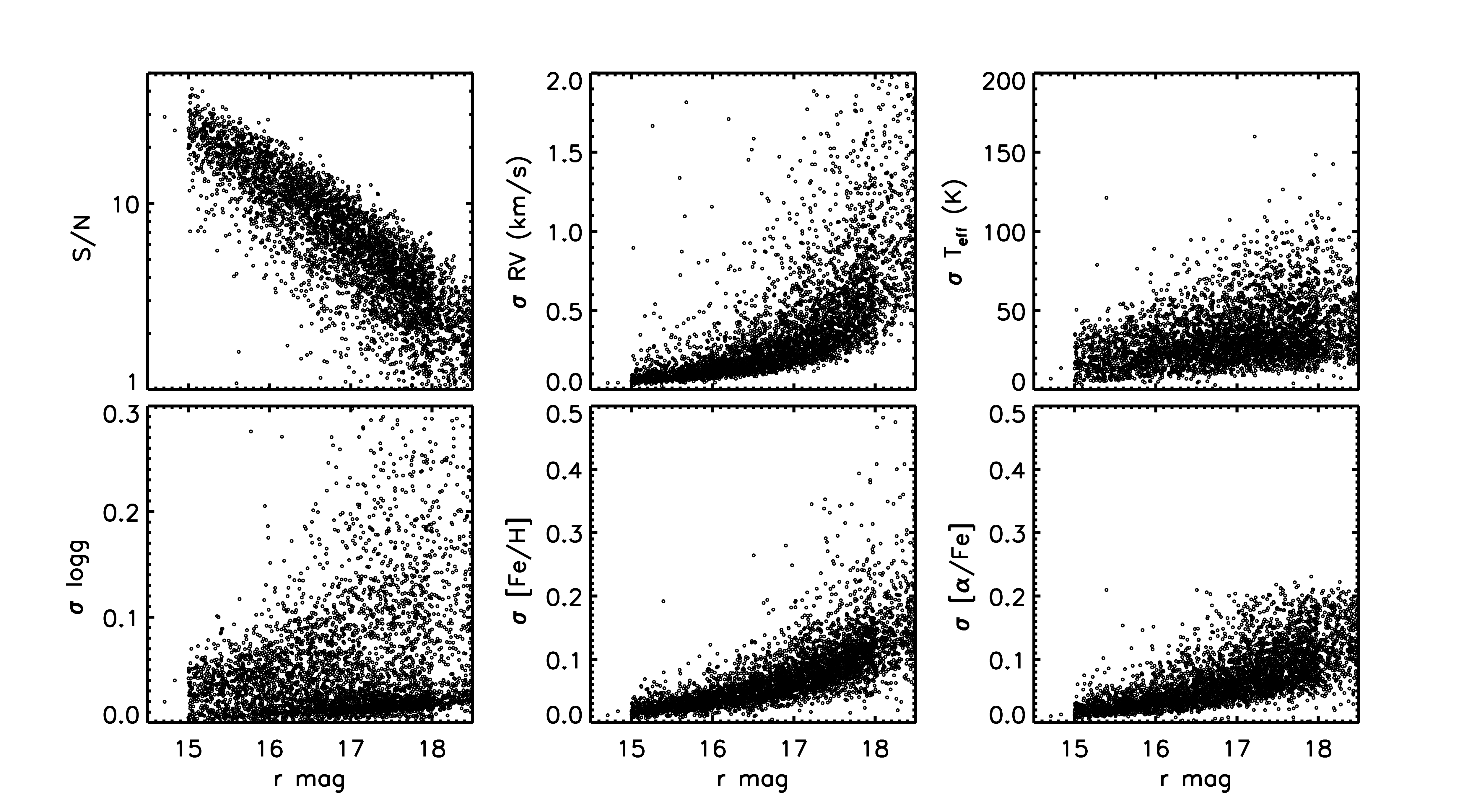}
\caption{Measurement uncertainties of derived parameters (and S/N)
  vs. $r$-band magnitude.  Only 5\% of the current sample is shown for
  clarity.  S/N is quoted per pixel.  The two sequences seen in
  $\logg$ panel reflect the different uncertainties obtained for
  dwarfs vs. giants (the former being smaller than the latter).}
\vspace{0.1cm}
\label{fig:uncert}
\end{figure*}

\subsection{Observations}

Data are obtained with the medium-resolution Hectochelle spectrograph
\citep{Szentgyorgyi11} on the MMT.  Hectochelle employs a robotic
fiber positioning system \citep{Fabricant05}, enabling the placement
of 240 fibers over a $1^{\circ}$ diameter field of view.  We use the
$110\,l \,mm^{-1}$ grating and the RV31 order-blocking filter.  The
data are binned by a factor of three in the wavelength direction and
two in the spatial direction.  This setup delivers a resolution of
$R\approx23,000$ over the (single order) wavelength range
$5150$\AA$-5300$\AA.  This corresponds to a velocity resolution of
$\sigma=5.5\kms$ \citep[see also][]{Walker15}.

In Figure \ref{fig:rv31} we show a detailed view of four spectral
types in order to highlight the atomic and molecular features in the
$5150$\AA$-5300$\AA\, wavelength range.  We show four spectra: the
Solar and Arcturus spectral atlases \citep[from][]{Hinkle00}, a model
M dwarf spectrum (computed from the same grid of models discussed in
Section \ref{s:ms}), and the spectrum of a well-studied metal-poor
giant HD122563, obtained through the UVES archive \citep{Bagnulo03}.
All of these spectra were convolved to a resolution of $R=23,000$.
For the Sun we also include the native resolution spectrum in order to
assess the line broadening due to the instrument resolution.  For the
model M dwarf we have computed a model without the MgH molecular lines
in order to highlight the overwhelming impact of this molecule for
metal-rich cool dwarfs in this spectral range.  Many of the strongest
features are labeled \citep[line identification is adopted
from][]{Hinkle00}.

The survey is being awarded time through the regular TAC processes at
the CfA and U. Arizona.  The allocations have resulted in
approximately 50 nights awarded to the survey per year.  Observations
are queue scheduled during bright time.  The nominal exposure time is
$3\times10$ min, although sub-optimal weather conditions occasionally
necessitate one or more additional exposures.  The exposure time is
set by the desire to observe fields as fast as possible while not
being dominated by overheads, which are $\approx15$ min per field.
This exposure time delivers S/N per pixel of $\approx2$ at the faint
limit of the main sample ($r=18$; see Figure \ref{fig:uncert} below).
Twilight flat fields are taken most nights and offer an important test
of the long-term stability of the derived RV measurements.  The
twilights are also used for relative fluxing (see below).

The actual number of fibers assigned to science targets varies in the
range of $170-200$, even when the total number of targets in the field
exceeds $400$.  This is due to fiber ``collisions'' resulting from the
fact that the fibers are robotically moved from the edge of the field
inward.

\vspace{1cm}

\subsection{Data reduction pipeline}

Data are reduced using HSRED
v2.1\footnote{https://bitbucket.org/saotdc/hsred}, an IDL-based
pipeline originally developed for Hectospec, but recently extended to
handle Hectochelle data as well (HSRED is based on the SDSS
\texttt{idlspec2d} algorithms). The code performs wavelength
calibration using a 5th order Legendre fit to the lines identified in
ThAr calibration spectra. Flat fielding is performed using a
combination of twilight sky flats and dome flats to simultaneously
solve for fiber-to-fiber throughput variations and the pixel-to-pixel
sensitivity. Dome flats are also used for identifying the traces of
each fiber on the chip, which are then extracted using an optimal
extraction algorithm. Cosmic rays are removed by an algorithm that
identifies outliers in a moving window across all science
exposures. Science exposures are sky subtracted by generating an
oversampled B-spline model of the sky from the $35-40$ designated sky
fibers in each observing configuration. Multiple exposures are
combined after extraction, using an inverse-variance weighting to
optimize signal-to-noise, though simple summed spectra are also
generated. Spectra are not interpolated onto a uniform, linear (or
log-linear) wavelength grid.  The spectra are not flux calibrated.
The first 120 fibers are assigned to one CCD while the second 120 are
assigned to a second CCD.

The relative system throughput as a function of wavelength is
estimated with the following procedure.  All available twilight
spectra are first divided by a continuum-normalized resolution-matched
spectrum of the Sun.  This removes small-scale features not due to
system throughput variations.  The resulting spectra are then median
combined.  The resulting stacked spectrum is fit with a 15th order
Chebyshev polynomial.  This procedure is performed for each of the two
CCDs.  This effective throughput curve is then used in the preparation
of the data for stellar parameter determination.

\subsection{\texttt{MINESweeper} stellar parameter pipeline}
\label{s:ms}

\subsubsection{Overview}

Stellar parameters are determined using \texttt{MINESweeper}
\citep{Cargile19}.  Briefly, the program jointly fits the Hectochelle
spectrum and the broadband photometric SED to a model that is based on
\texttt{MIST} \citep{Choi16} stellar isochrones.  The position along
the isochrone is determined by the stellar mass, age, and metallicity.
At each point along the isochrone there is a corresponding model SED
and high resolution spectrum.  We include photometry from Pan-STARRS,
SDSS, {\it Gaia}, 2MASS, and {\it WISE} where available.

The spectral models (and corresponding photometry) are computed from
grids of model atmospheres and the spectrum synthesis code
\texttt{SYNTHE} \citep{Kurucz93}.  The model atmospheres are computed
with the \texttt{ATLAS12} program \citep{Kurucz70}.  Both atmospheres
and spectra are computed in 1D assuming plane-parallel geometry and
LTE.  We adopt the solar abundances from \citet{Asplund09}, which is
also the abundance scale used in the \texttt{MIST} isochrones.  For
the spectral synthesis, a constant microturbulence of $v_t=1\kms$ is
adopted for all spectra.  The spectral grids include variable
[$\alpha$/Fe] abundances, though we note that the isochrone tables do
not currently include this level of flexibility.  Atomic and molecular
line lists are adopted from the latest compilation of R. Kurucz, and
have been astrophysically-calibrated against ultra high resolution
spectra of the Sun and Arcturus using the same model assumptions as
adopted herein.

In addition to the mass, age, and metallicity that determine the
location along an isochrone, \texttt{MINESweeper} also determines the
distance, reddening ($A_{\rm V}$), [$\alpha$/Fe], radial velocity
(RV), stellar broadening (representing both rotational broadening and
macroturbulence), instrumental resolution, and four Chebyshev
coefficients for matching the spectral shape between the data and
model.  Recall that the initial preparation of the data includes a
continuum normalization procedure based on the twilight exposures.
Experimentation led us to settle on a 4th order Chebyshev polynomial
for fine-tuning the continuum matching between the data and model.
The total number of free parameters being estimated from the data is
therefore 13.

Each free parameter requires a prior.  Since we are focused on the
stellar halo, which is comprised of old stars, we adopt a uniform log
age prior of [10.0,10.15] log(yr).  This is consistent with the
approach taken by many other authors \citep[e.g.,][]{Burnett10, Xue14,
  Schonrich14}.  A \citet{Kroupa01} IMF is adopted for the mass prior.
A first pass estimate of the RV is determined by a simple
cross-correlation using a template with $\teff$ determined from a
first-pass fit to the SED and fixed values of [Fe/H]$=-1.0$,
[$\alpha$/Fe]$=0.0$, and $\logg=2.0$.  The prior on RV in the main
program is taken to be uniform centered on the initial guess with a
width of $\pm25\kms$.  The reddening prior is a Gaussian centered on
the \citet{Schlegel98} dust map value with a width of 15\% \citep[we
have adopted the 14\% lower normalization of the dust maps provided
by][]{Schlafly11}.  The prior on metallicity is uniform in the range
[$-4.0$,0.5], and [$\alpha$/Fe] from [$-0.2$,0.6].  Rotational
broadening is restricted to [0,15] $\kms$ with an additional Gaussian
prior centered on zero with a width of $3\kms$.  The prior on the
instrumental resolution is a Gaussian centered on $R=25,000$ with a
standard deviation of $1,000$.  {\it Gaia} DR2 parallaxes are used as
a Gaussian prior, which aides in the separation between dwarf and
giant solutions even when the {\it Gaia} parallax has low S/N.  We
adopt a {\it Gaia} parallax zeropoint offset of $+0.05$, consistent
with recent literature estimates \citep[e.g.,][]{Leung19,
  Schonrich19}.  Sampling is performed in log distance, which is
equivalent to assuming a prior on the stellar density profile of
$n\propto d^{-3}$.  See \citet{Bailer-Jones18} for an example of a more
complex set of priors for determining distances.  Finally, we note
that the isochrones are restricted to evolutionary points between the
zero-age main sequence and the beginning of the first thermal pulse.

\begin{figure*}[!t]
\center
\includegraphics[width=1.\textwidth]{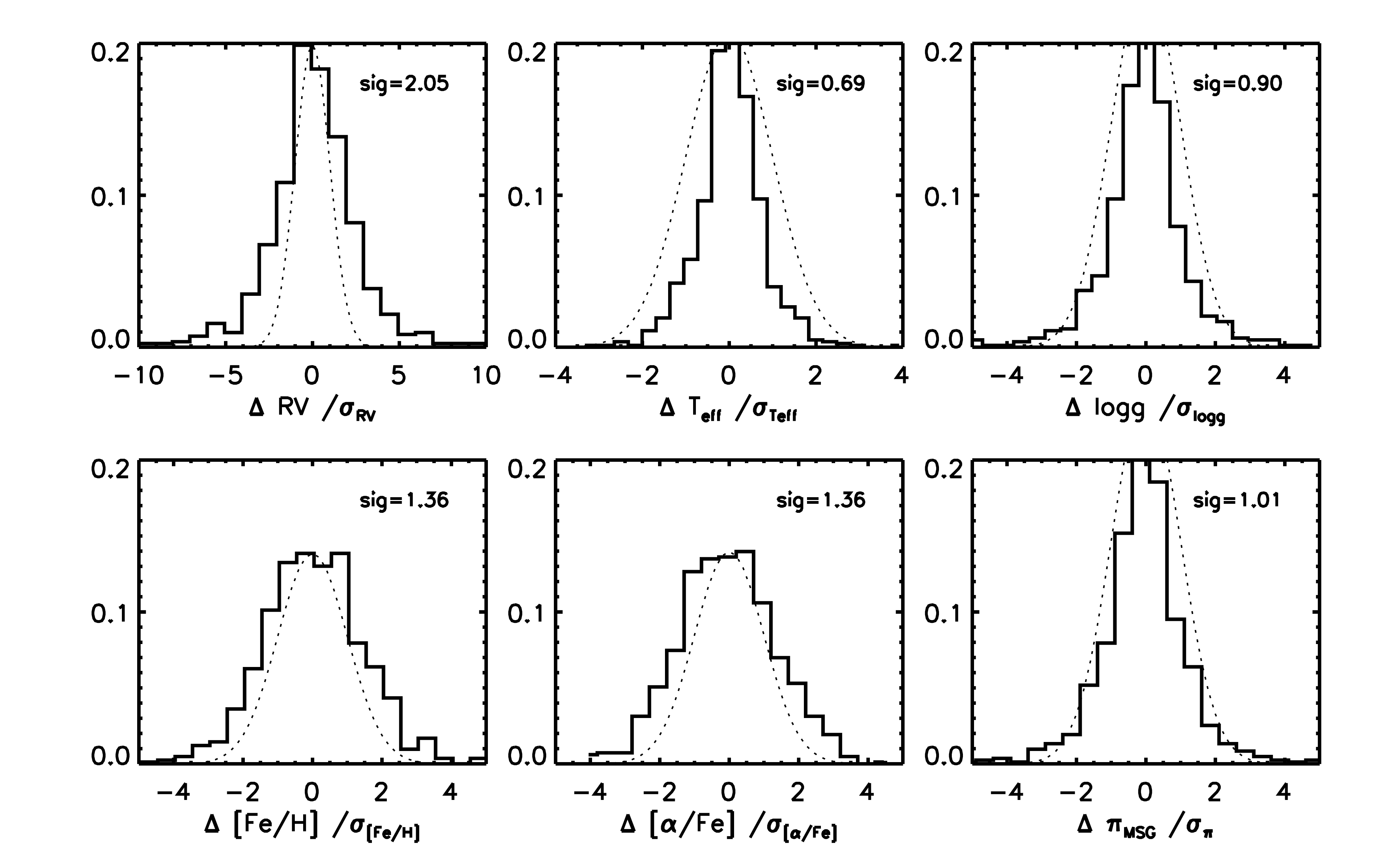}
\caption{Difference in best-fit parameters for stars observed on two
  occasions.  Differences are quoted in units of the quadrature-summed
  measurement uncertainties.  There are approximately 900 duplicates
  in the sample.  Also included in each panel is the 3-sigma clipped
  standard deviation of the distribution and a unit Gaussian
  distribution (dotted lines).  In general the values are near one,
  indicating that the pipeline is reporting reliable uncertainties.
  The one exception are the RVs, where the formal uncertainties appear
  to be under-estimated by a factor of $\approx2$ (note that the
  median formal uncertainty on the RVs is $0.24\kms$).}
\label{fig:dup}
\end{figure*}

\begin{figure*}
\center
\includegraphics[width=0.90\textwidth]{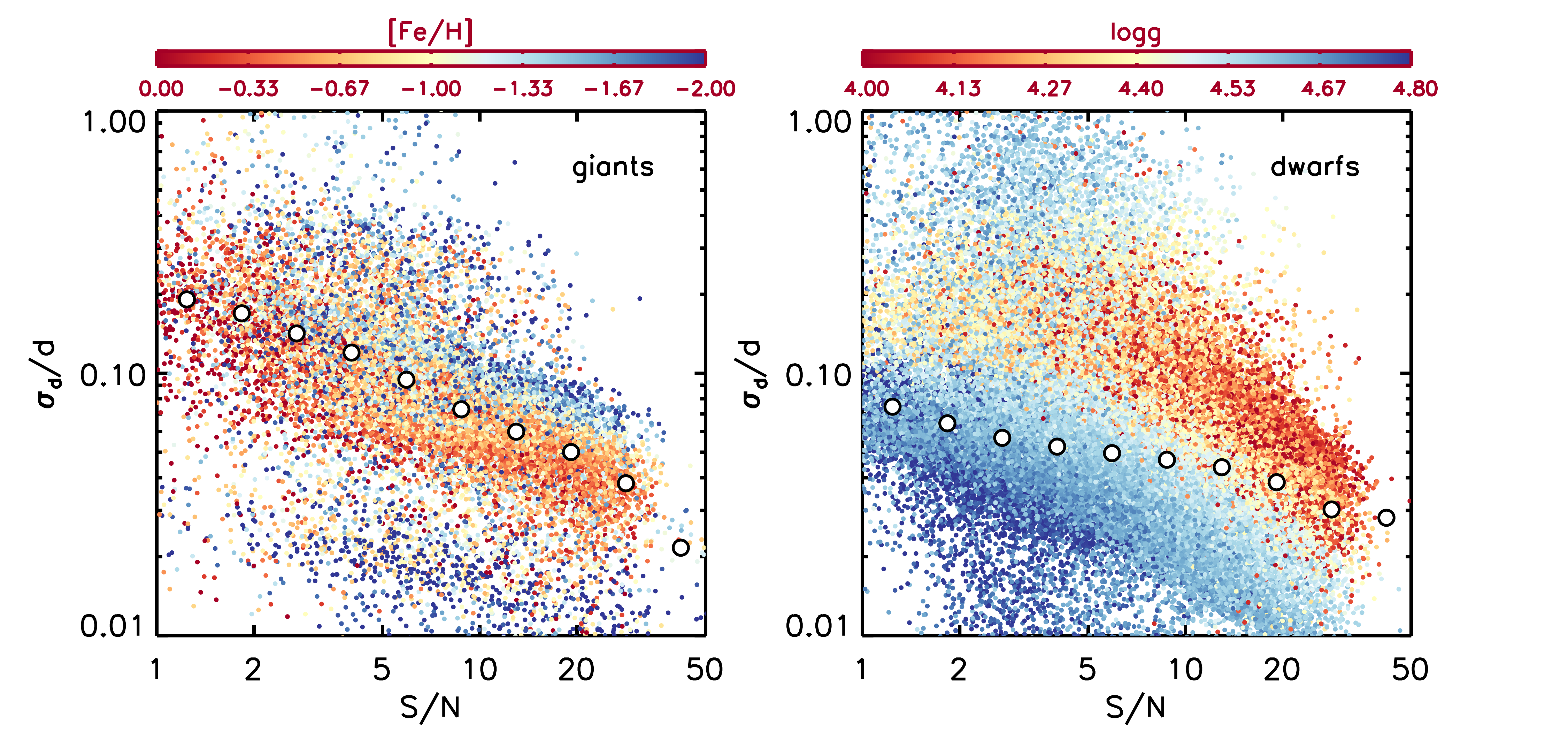}
\vspace{0.1cm}
\caption{Fractional distance uncertainties vs. S/N.  Median values are
  shown as large open symbols.  Left Panel: Results for giants
  ($\logg<4.0$), color-coded by metallicity.  At fixed S/N, the
  fraction distance uncertainties are larger at lower metallicity.
  Right Panel: Results for dwarfs ($\logg>4.0$), color-coded by
  $\logg$.}
\label{fig:dist}
\end{figure*}

\begin{figure*}[!t]
\center
\includegraphics[width=0.85\textwidth]{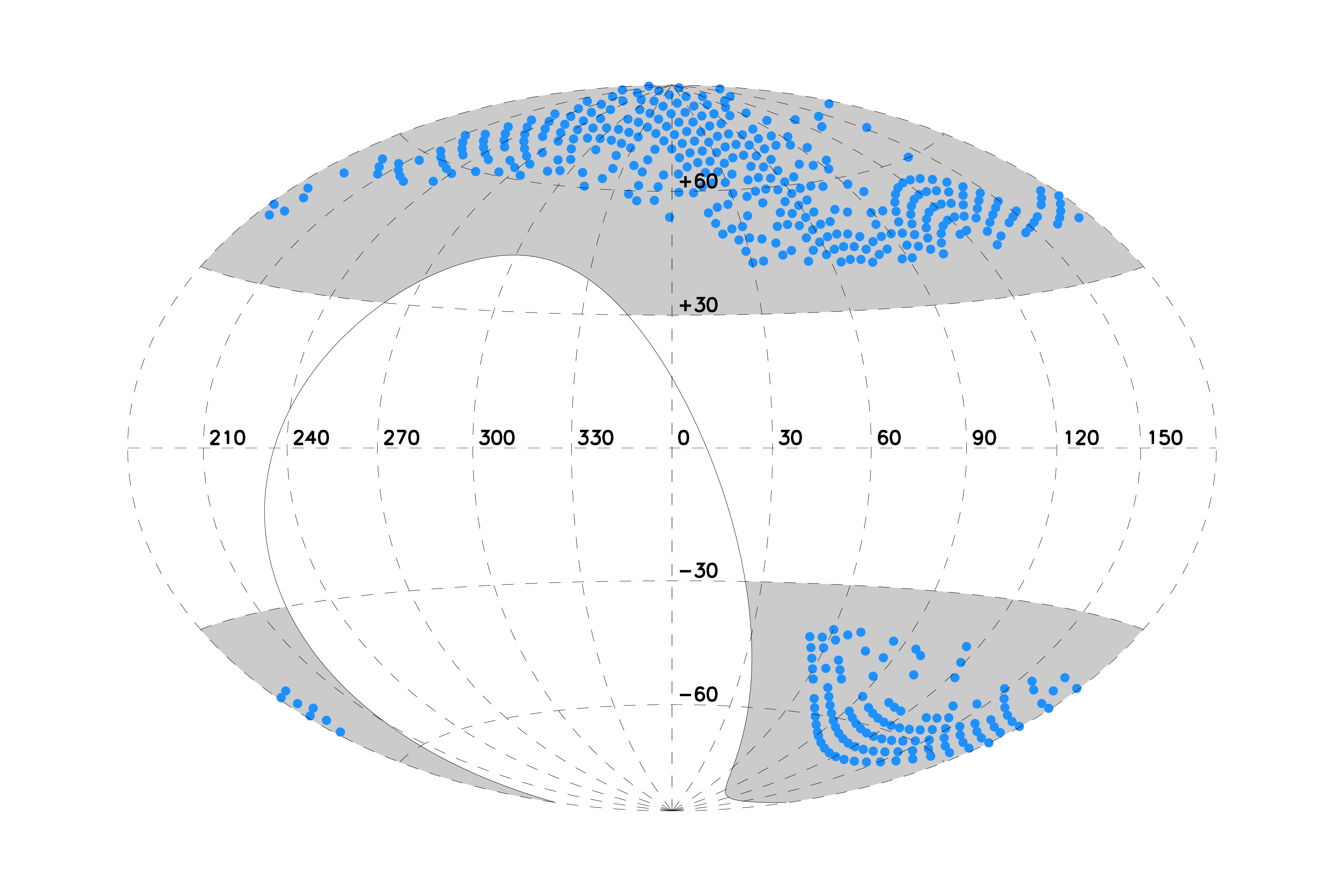}
\caption{Survey footprint (grey) in Galactic coordinates.  Fields
  observed as of June, 2019 are shown as blue symbols (not drawn to
  scale).  The solid line traces Dec.$=-20^\circ$.}
\label{fig:footprint}
\end{figure*}

\begin{figure*}[!t]
\center
\includegraphics[width=0.85\textwidth]{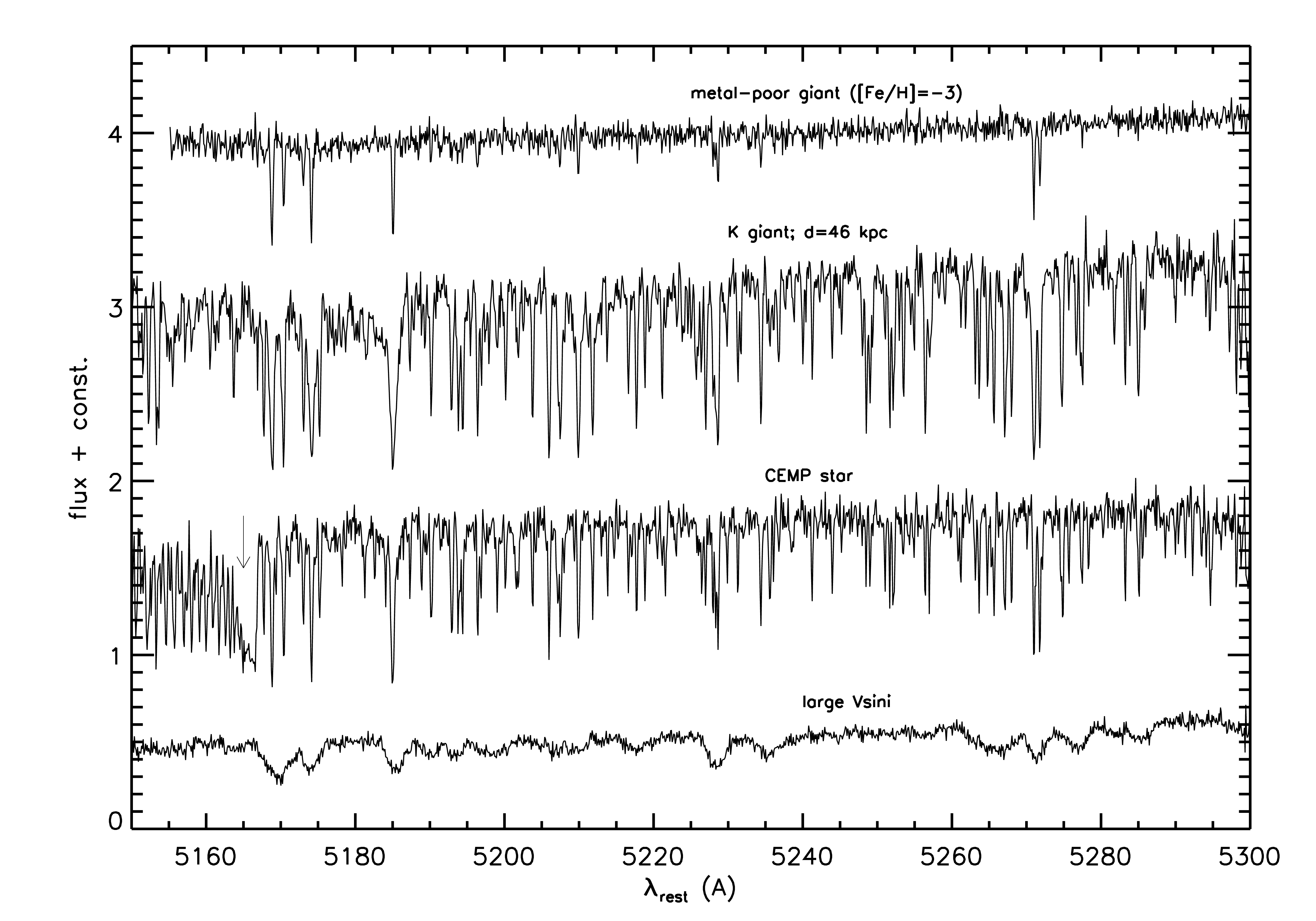}
\vspace{0.1cm}
\caption{Several interesting and unusual spectra.  From top to bottom:
  a metal-poor giant with [Fe/H]$=-3.0$, $\logg=1.5$, and a distance
  of 25 kpc; a star selected with our ``K-giant'' color-cuts at a
  distance of 46 kpc and [Fe/H]$=-1.0$; a carbon-enhanced giant at a
  preliminary distance of 19 kpc (the C$_2$ molecular feature is
  indicated with an arrow); a star with large projected rotation
  velocity.  The S/N of these spectra are $\gtrsim20$.}
\label{fig:specdemo}
\end{figure*}

\begin{figure*}[!t]
\center
\includegraphics[width=0.9\textwidth]{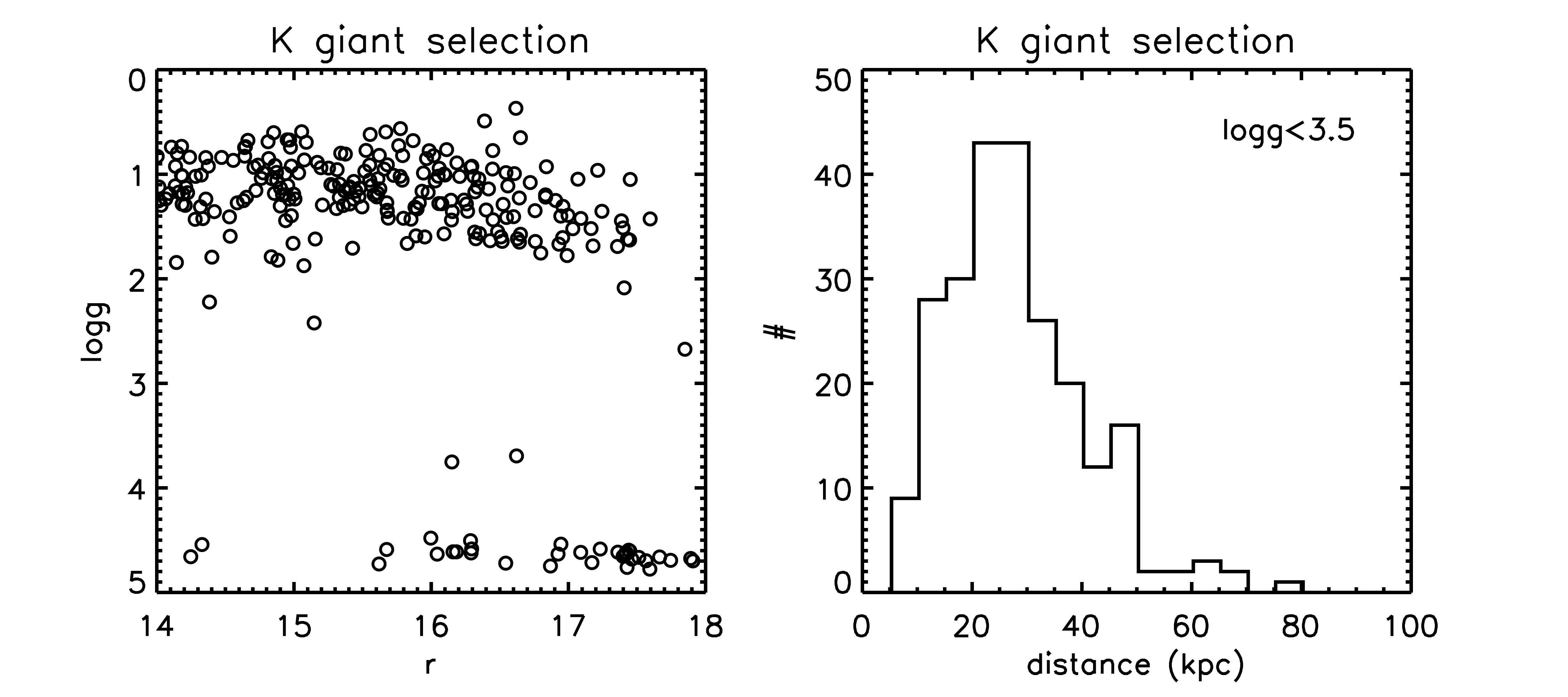}
\vspace{0.1cm}
\caption{Validation of the photometric K-giant selection from
  \citet{Conroy18a}.  Left Panel: $\logg$ vs. $r-$band magnitude for
  the 272 photometrically-selected K-giants observed to-date with
  S/N$>2$.  Giants comprise 87\% of the sample.  Right Panel:
  distribution of distances for the giants.  The median distance is 24
  kpc with the most distant star at 79 kpc.}
\label{fig:kgiant}
\end{figure*}

\begin{figure}
\center
\includegraphics[width=0.48\textwidth]{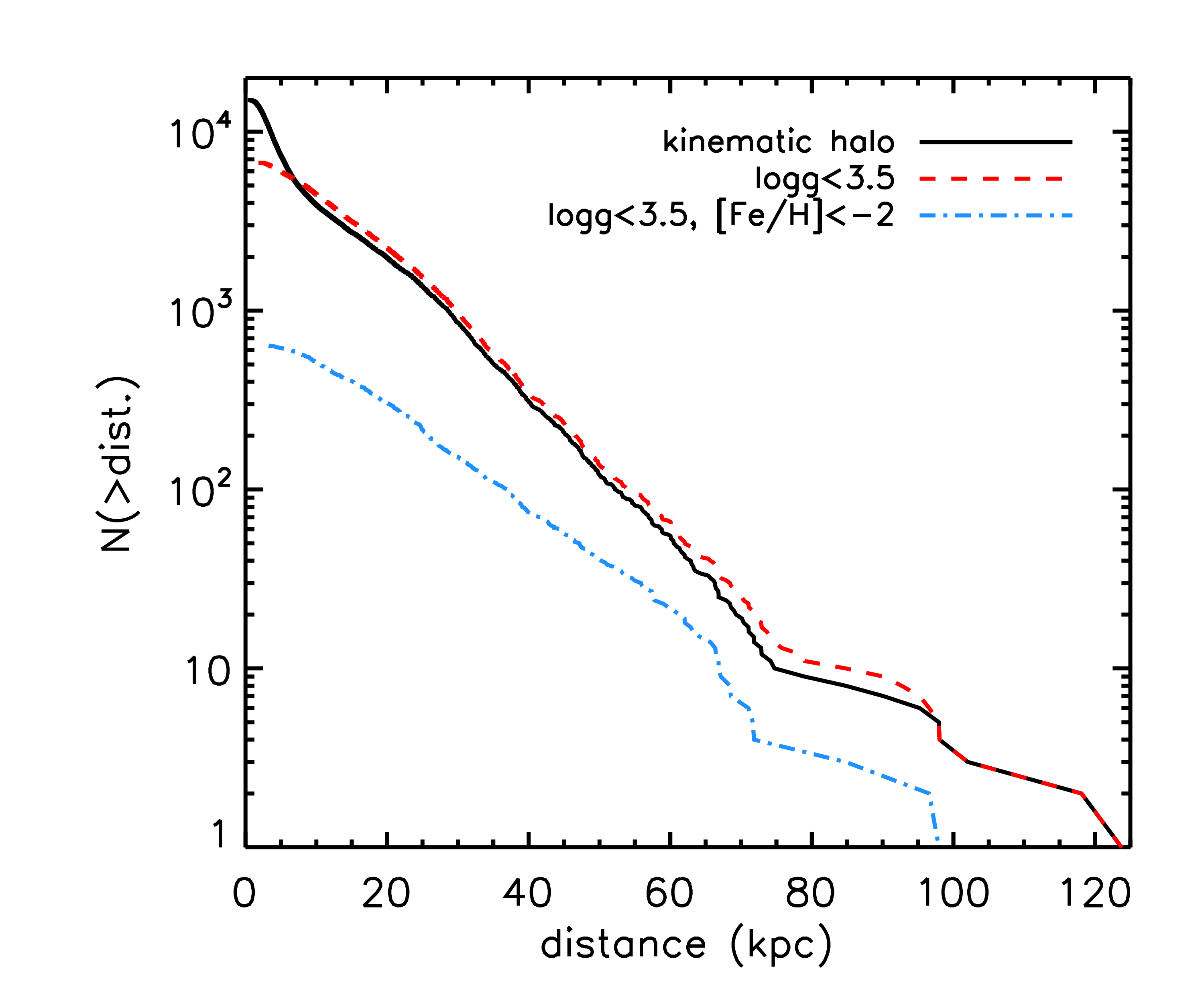}
\caption{Cumulative histogram of heliocentric distances to stars in
  the H3 survey.  Spectrophotometric distances are estimated via the
  \texttt{MINESweeper} program.  We show the distribution of
  kinematically-selected halo stars (solid line), giants ($\logg<3.5$;
  dashed line), and low-metallicity giants ($\logg<3.5$ and
  [Fe/H]$<-2$; dot-dashed line).  Results are shown for data collected
  as of June, 2019.}
\label{fig:dhist}
\end{figure}

\begin{figure*}[!t]
\center
\includegraphics[width=0.95\textwidth]{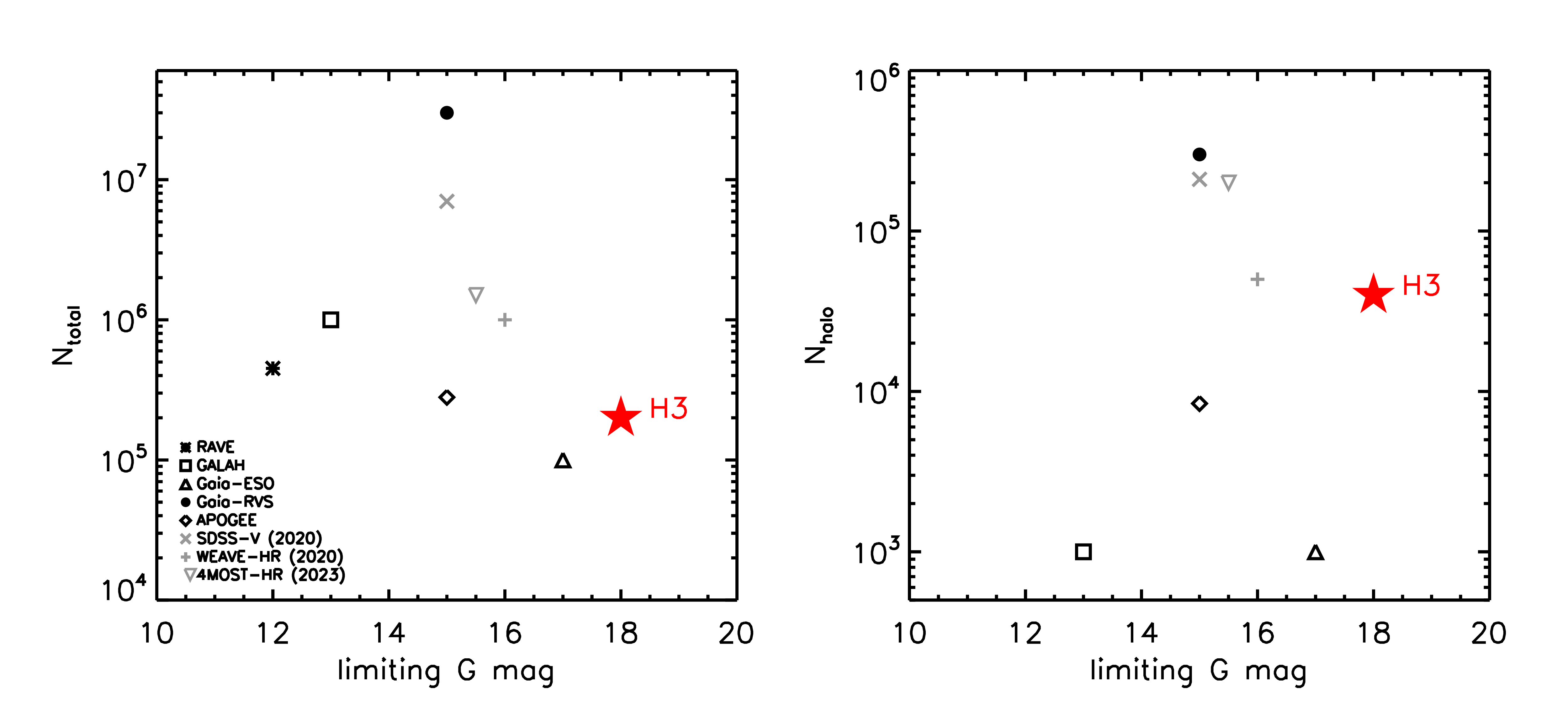}
\vspace{0.1cm}
\caption{Comparison of existing and planned stellar spectroscopic
  surveys with a spectral resolution of $R>8,000$.  Left Panel: total
  number of stars expected as a function of the limiting G-band
  magnitude.  Right Panel: Estimated number of {\it halo} stars from
  each survey (see text for details).  Ongoing or completed surveys
  are shown as black points, while upcoming surveys are shown as grey
  (along with anticipated survey start dates in parentheses).}
\label{fig:comp}
\end{figure*}

\texttt{MINESweeper}\, enables measurements of metallicities to values
as low as $-4.0$, although we caution that estimates below
$\approx-2.3$ have not been validated against literature estimates
(see the next section).  The current setup enables measurement only of
[Fe/H] and [$\alpha$/Fe].  However, in the wavelength range of the
survey there are atomic and molecular transitions from elements
including C, Na, Mg, Ca, Ti, V, Cr, Mn, Fe, Co, Ni, Cu, Y, Ce, and Nd.
Many of these transitions are relatively weak (line depths of a few
percent) and so will be impossible to measure for the majority of the
low SNR spectra in our sample.  However, for the bright subset, we
intend to apply complementary techniques \citep{Ting19} in order to
measure at least some of these valuable element abundances.

The 13 free parameters are fit to the data using the \texttt{dynesty}
nested sampling package \citep{Speagle19}.  \texttt{MINESweeper}
currently takes $1-2$ CPU hr per star to fully sample the posterior
space.

\vspace{2cm}

\subsubsection{Pipeline validation}

At the beginning of each observing night twilight spectra are obtained
to aid in the data reduction.  We have fit all of these spectra with
\texttt{MINESweeper} in order to test the absolute accuracy of the
radial velocities (the twilight spectra are dominated by Solar
absorption lines).  Taking into account the gravitational redshift and
convective blueshift of the Solar spectrum, \texttt{MINESweeper}
recovers absolute velocities with an accuracy of $\pm0.5\kms$.

\citet{Cargile19} fit Hectochelle spectra obtained for several open
and globular clusters and showed that \texttt{MINESweeper} accurately
recovers metallicities, $\alpha$-enhancements, distances, and
locations in $\logg-\teff$ space for clusters over a wide range of
metallicities (from M92 at [Fe/H]$=-2.3$ to M67 at [Fe/H]=$0.0$).
There are modest systematic uncertainties as a function of $\logg$, at
the $\approx0.1$ dex level.  Cargile et al. also show comparisons
between spectrophotometric distances and {\it Gaia} parallaxes for H3
stars with high S/N {\it Gaia} parallaxes.  For this comparison these
authors did not use the {\it Gaia} parallaxes as a prior in the
fitting.  They found very good agreement with {\it Gaia} parallaxes.

Figure \ref{fig:uncert} shows the distribution of S/N and
uncertainties in RV, $\teff$, $\logg$, [Fe/H], and [$\alpha$/Fe] as a
function of $r-$band magnitude.  We show here only 5\% of the current
sample for clarity.  At the faint end of the primary sample ($r=18$),
the RV uncertainties are $<1\kms$ and abundances are precise to
$\approx0.1$ dex.  Temperatures are measured to $\approx50$ K.  This
level of precision is comparable to standard color$-\teff$ relations
\cite[e.g.,][]{Casagrande10}.  The $\logg$ uncertainties display two
sequences, one for dwarfs (with smaller uncertainties) and the other
for giants (with larger uncertainties).  This is a natural consequence
of the behavior of isochrones in $\teff-\logg$ space: a given
measurement precision in $\teff$ corresponds to a much narrower range
of allowed $\logg$ values on the main sequence compared to the giant
branch.

In the course of the survey 12 fields were observed twice, once with
the color-based selection, and once with the parallax-based selection.
This allowed us to assess the impact of the target selection (e.g.,
Figure \ref{fig:hfrac}).  As a consequence of this duplication,
$\approx1,000$ stars were observed twice, which provides an opportunity
to test the fidelity of the stellar parameter pipeline.  Figure
\ref{fig:dup} shows the difference in best-fit parameters for the
duplicate observations, plotted in units of the parameter
uncertainties.  In general the differences follow a Gaussian with a
3-sigma clipped standard deviation close to one.  This means that the
pipeline uncertainties are reliable.

The panel comparing RVs deserves further comment.  First, the pipeline
uncertainties appear to be under-estimated by a factor of two.  This
is of little concern for our science applications, as the quoted
uncertainties are very small (the median uncertainty is $0.24\kms$;
see Figure \ref{fig:uncert}).  As noted earlier, analysis of the
twilight spectra indicates an absolute uncertainty floor of
$\approx0.5\kms$ in the RVs.  Adding this uncertainty floor in
quadrature to the statistical uncertainties results in a total error
budget in good agreement with the repeat observations.  Second, there
is a small but prominent population of outliers.  It appears that many
of these outliers are due to binarity.  In some cases a second pair of
lines is clearly visible in the spectrum, while in other cases the
lines appear significantly broadened (e.g., by $>5\kms$).  These
outliers will provide a valuable probe of the impact of binarity on
our single-epoch RV estimates.

Figure \ref{fig:dist} shows the fractional distance uncertainties
derived from \texttt{MINESweeper}.  Results are shown separately for
giants and dwarfs.  In each panel the results are color-coded by the
variable that most strongly correlates with the distance uncertainties
at fixed S/N ([Fe/H] for the giants and $\logg$ for the dwarfs).  For
the dwarfs, the median uncertainties are $5$\% for S/N$>2$, while for
the giants the uncertainties are $8$\% for S/N$>2$.  These
uncertainties are broadly consistent with previous work
\citep[e.g.,][]{Xue14, Wang16}.  See Cargile et al. for further
discussion of the \texttt{MINESweeper} distance estimates.


\subsubsection{Areas for improvement}

The current pipeline performs very well, as detailed in Cargile et
al. and summarized above.  Nonetheless, there are several areas where
improvement is warranted.  First, the current isochrone tables do not
allow for variation in [$\alpha$/Fe].  This limitation is currently
being addressed within the \texttt{MIST} collaboration.  Second, the
lower main sequence of standard stellar models does not reproduce the
observations \citep[e.g.,][]{Feiden12, Choi16}.  The reason for this
discrepancy is believed to be related to magnetic fields, which
standard stellar models do not include.  We are currently working on a
simple and empirical starspot model to add on top of the \texttt{MIST}
isochrones, which will by construction result in better agreement with
observations at the lower main sequence.  We are also considering more
informative priors on the distances \citep[as in
e.g.,][]{Bailer-Jones18}.

Stellar binarity is an important source of systematic uncertainty in
the current pipeline.  Binarity will influence our data in two
possible ways: as double-lined spectroscopic binaries, and as
unresolved photometric binaries.  Both will bias the derived stellar
parameters in systematic ways, though we expect this effect to be
limited to dwarfs, as giants will greatly outshine a dwarf companion.
We are developing techniques for identifying binaries, but this will
remain a challenge especially when the {\it Gaia} parallaxes are of
low S/N and there is no obvious spectroscopic signature of a binary.

On the spectral modeling side, we plan to re-visit the astrophysical
calibration of the line list used to generate the synthetic spectra.
In previous work the line list was tuned only to the Sun and Arcturus.
By adding several additional calibrators (e.g., Barnard's star) we
hope to achieve higher accuracy for the coolest stars.  Finally, we
have fixed the microturbulent velocity, $v_t$, to $1\kms$.  However,
this parameter is known to vary with stellar type and evolutionary
phase \citep[e.g.,][]{Ramirez13}, so in subsequent work we intend to
expand the spectral grid so that $v_t$ can be a fitted parameter.


\section{Progress To-Date}
\label{s:prog}

The survey began collecting data in the Fall of 2017.  As of June, 2019
we have observed 469 fields and collected 87,930 spectra of 86,598
stars.  The locations of the observed fields in Galactic coordinates
are shown in Figure \ref{fig:footprint}.  

In the remainder of this section we provide a brief overview of the
survey data.  Figure \ref{fig:dhist} shows the distribution of
heliocentric distances of kinematically defined halo stars, giants,
and metal-poor giants.  Note that this is simply the cumulative count
distribution and does not include a correction for the selection
function.  An estimation of the halo density profile from these data
will be the subject of future work.

Figure \ref{fig:specdemo} shows a gallery of several interesting and
unusual spectra from the current sample.  The spectra include (from
top to bottom): a metal-poor giant with [Fe/H]$=-3.0$, $\logg=1.5$,
and a distance of 25 kpc; a star selected with our ``K-giant''
color-cuts at a distance of 46 kpc; a carbon-enhanced giant at a
preliminary distance of 19 kpc (the C$_2$ molecular feature is
indicated with an arrow).  Carbon-enhanced metal-poor (CEMP) stars are
known to be common at low metallicity, with an estimated frequency
$>10$\% at [Fe/H]$<-2$ \citep{Lee13}.

The last spectrum in Figure \ref{fig:specdemo} is a star with a large
projected rotation velocity.  While unusual, this type of star is not
unique in our survey: there are several hundred stars that have
pipeline-based rotation velocities that are limited by the imposed
prior of $15\kms$.  By-eye inspection of these stars indicates that
many have much higher rotational velocities.  Some of these stars have
high S/N {\it Gaia} parallaxes and so can be placed on an
absolute-magnitude CMD.  Doing so reveals that most of these stars
reside on or near the equal mass binary star sequence.  It will be
interesting to follow up these stars in detail as a unique probe of
the binary star population in the stellar halo.

As discussed in Section \ref{s:sel}, we have included in our target
selection the photometrically identified K giants from
\citet{Conroy18a}.  These stars are rare but of potentially high value
as they can be seen to great distances ($>100$ kpc).  In Figure
\ref{fig:kgiant} we show the spectroscopically confirmed $\logg$
values for this sample as a function of $r-$band magnitude.  The
measured $\logg$ values confirm the results presented in
\citet{Conroy18a} that the photometric K giant selection is highly
pure - in our survey 87\% of these stars are bone fide giants.  The
purity decreases at fainter magnitudes, which we believe is due to
increasing photometric scatter in the {\it WISE} photometry used in
the selection criteria. The right panel shows the distribution of
distances of these stars; the mean distance of 24 kpc and the most
distant star is at 79 kpc.

\begin{figure*}[!t]
\center
\includegraphics[width=0.95\textwidth]{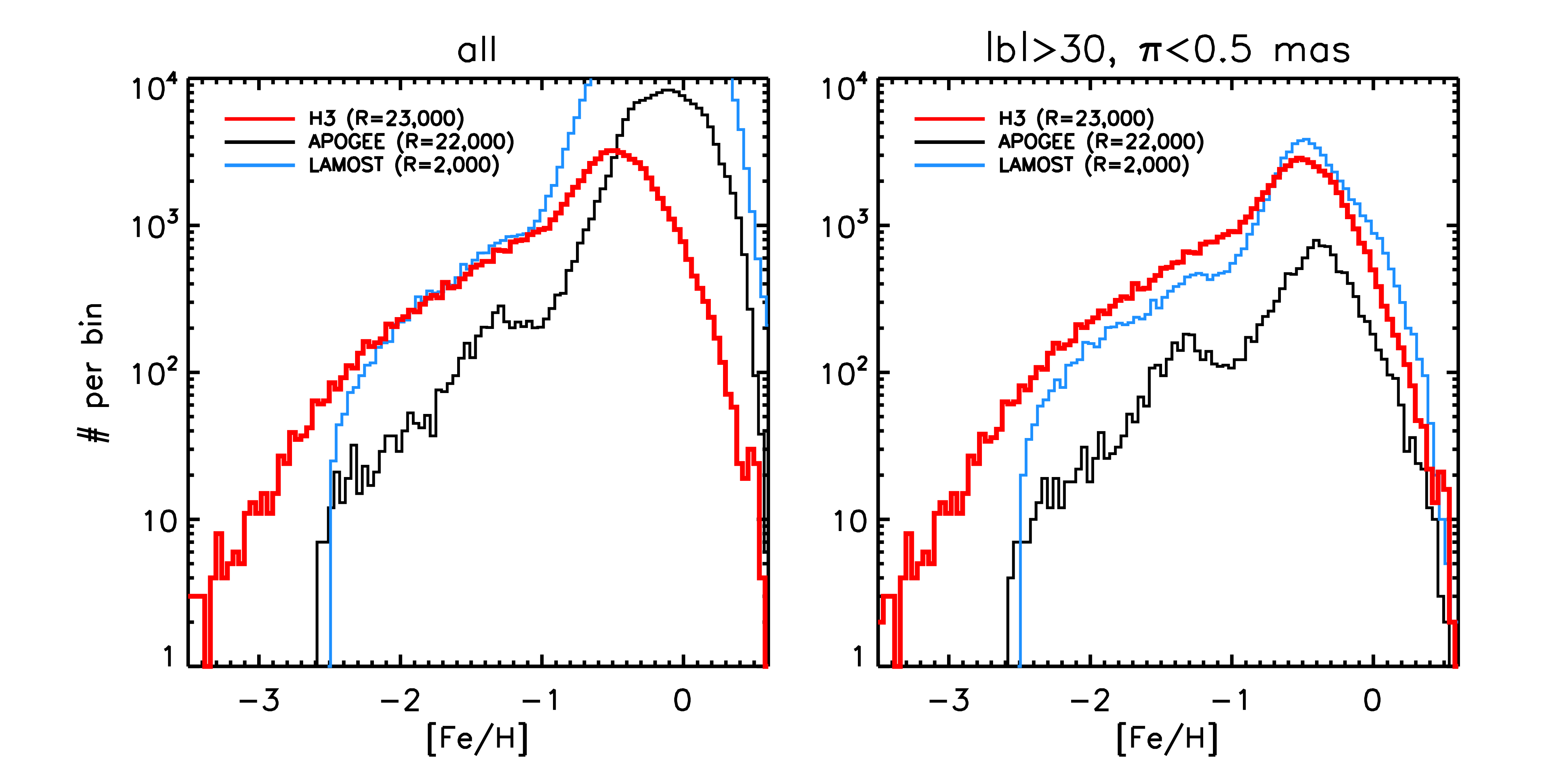}
\vspace{0.1cm}
\caption{Left Panel: Metallicity distribution of the current H3 Survey
  compared to the APOGEE and LAMOST Surveys.  Right Panel: Metallicity
  distributions for the three surveys for all stars with Galactic
  latitude $|b|>30^\circ$ and a Gaia DR2 parallax $\pi<0.5$ mas.  H3
  stars were selected according to these criteria, whereas the other
  two surveys were not.  Comparison between the left and right panels
  highlights the fact that while H3 is not the largest spectroscopic
  survey, it is by far the largest medium-resolution survey of distant
  high latitude stars, and rivals low-resolution surveys of such
  populations.  Histograms were computed with a bin size of 0.04 dex.}
\label{fig:feh}
\end{figure*}


\section{H3 in context}
\label{s:context}

There are multiple recently completed, ongoing, and planned
large-scale ground-based stellar spectroscopic surveys.  In this
section we place H3 in context with these other surveys.

Figure \ref{fig:comp} compares a number of large-area
medium-resolution ($R>8,000$) spectroscopic surveys. The left panel
shows the total number of stars from each survey as a function of the
limiting $G-$band magnitude.  The surveys include the completed RAVE
\citep{Steinmetz06} and Gaia-ESO \citep{Gilmore12} surveys, the
ongoing H3, APOGEE \citep{Majewski17} and GALAH \citep{Martell17}
surveys, and the upcoming SDSS-V \citep{Kollmeier17}, WEAVE
\citep{Dalton12}, and 4MOST \citep{Guiglion19} surveys.  In the case
of ongoing or planned surveys we plot their quoted final number of
expected targets.  Most surveys have limiting magnitudes in filters
other than the $G-$band (e.g., APOGEE and SDSS-V are $H-$band
selected), so the translation to a limiting $G-$band magnitude is
approximate.  The {\it Gaia} G-band is close to the SDSS and
Pan-STARRS $r$-band (to within $\lesssim0.05$ mag) for
$4000<\teff<6500$ K.

We have not included low resolution stellar spectroscopic surveys,
such as the completed SEGUE survey \citep{Yanny09}, the ongoing LAMOST
survey \citep{Deng12}, and the upcoming DESI survey \citep{DESI}.
Moreover, while both of the upcoming WEAVE and 4MOST surveys have
medium and low resolution components; in Figure \ref{fig:comp} we only
include the medium resolution surveys.

We caution that it is not obvious whether medium or low resolution is
``better'' in the context of a fixed survey speed (e.g., number of
stars per night) and measurement precision requirements.  As argued in
\citet{Ting17}, holding the exposure time and the number of pixels per
spectrum fixed, the theoretical information content is nearly
independent of spectral resolution.  Of course, at low spectral
resolution one has to contend with a greater variety of
``sub-resolution'' effects, including line blending.  For example,
$R=2,000$ corresponds to a velocity dispersion of $63\kms$, so
velocity precision of $<5\kms$ (likely the relevant scale for
identifying substructure in phase space) requires identifying line
centers at the $1/10$ pixel level.  This is not impossible, and with
sufficient calibrating data should be achievable.  An argument in
favor of lower resolution is access to a larger number of transitions
from more species than would be available from a high resolution
spectrum, if the total number of pixels is held fixed.

\begin{deluxetable}{lr}
\tablecaption{H3 at a Glance}
\tablehead{ \colhead{} & \colhead{} }
\startdata
Telescope & MMT (6.5m; Mt Hopkins, AZ) \\
Instrument & Hectochelle \\
Spectral Resolution & $R=23,000$ \\
Wavelength Range & 5150\AA$-5300$\AA \\
Number of Stars & $N=200,000$ \\
Time-frame & $2017-2021$ \\
Magnitude Range & $15<r<18$ \\
Main Sample Selection & $\pi<0.5$ mas \\
Footprint & $|b|>30^{\circ}$; Dec.$>-20^{\circ}$ \\
Median S/N & 6 per pixel \\
Kinematic halo fraction & $\approx20$\% 
\enddata
\vspace{0.1cm} 
\label{t:h3}
\end{deluxetable}

Another important distinguishing feature in the landscape of surveys
is the hemispheric coverage.  Surveys covering the Northern hemisphere
are WEAVE, LAMOST, DESI, and H3.  Surveys covering the South are RAVE,
GALAH, Gaia-ESO, and 4MOST.  All-sky surveys include {\it Gaia},
APOGEE, and SDSS-V (the latter two have twin spectrographs operating
at Northern and Southern observatories).

It is challenging to provide comparisons across surveys with such a
wide variety of survey strategies.  In addition to depth, number of
stars, and spectral resolution, one must also consider exposure time
and hence S/N, wavelength coverage, targeting strategy, and on-sky
footprint (e.g., Northern vs. Southern hemispheres, disk vs. halo
fields, sparse vs. dense tiling).  Comparisons such as those shown in
Figure \ref{fig:comp} should therefore be interpreted with these
considerations in mind.

The right panel of Figure \ref{fig:comp} shows an estimate of the
number of halo stars that each survey has or will yield.  To construct
this plot we have assumed the following halo fractions: 0.1\% for
GALAH \citep[from][]{Martell17}, 1\% for Gaia-RVS \citep[estimated
from the][mock catalog]{Rybizki18}, 1\% for Gaia-ESO Survey
(G. Gilmore priv. comm.), 3\% for APOGEE and SDSS-V (this is the
kinematic halo fraction estimated from an APOGEE-{\it Gaia}
cross-matched catalog; we adopt the same fraction for SDSS-V), and
20\% for H3.  The numbers for WEAVE-HR are from Jin, S., et al. (in
prep.), and the 4MOST-HR halo population is from \citet{Christlieb19}.
The halo fraction in H3 is an order of magnitude higher than in any
other moderate-resolution survey. Among the surveys with an
appreciable number of halo stars, H3 is two magnitudes deeper.  For a
given stellar type (e.g., TRGB, red clump), this increases the maximum
observable distance by a factor of 2.5.

In Figure \ref{fig:feh} we provide a more detailed comparison between
three surveys: H3, APOGEE, and LAMOST.  In the left panel we show
histograms of metallicity for the entire sample from each survey,
subject to a few quality cuts (H3: S/N$>2$ and quality flag$=0$;
APOGEE: \texttt{ASPCAPFLAG}$=0$; LAMOST: S/N$_g>40$, and
\texttt{objtype}=`star').  LAMOST clearly dominates the overall
sample.  Notice that H3 has more low-metallicity ([Fe/H]$<-1$) stars
than APOGEE.  For APOGEE and LAMOST the metallicities are truncated at
$-2.5$, which is an artefact of the spectral grids used in estimating
stellar parameters.  Moreover, while the H3 metallicities have been
well-tested down to $-2.5$, we caution that the lowest metallicities
will require careful vetting.

The right panel of Figure \ref{fig:feh} shows the same histograms of
metallicity with the additional requirements that $|b|>30^{\circ}$ and
$\pi-2\sigma_{\pi}<0.5$ mas.  This is the selection function for H3,
so it is not too surprising that with these restrictions H3 contains
more metal-poor stars even than LAMOST.  Nonetheless, this plot
highlights the strengths of the H3 survey in comparison to existing
large spectroscopic surveys.  Comparison between the left and right
figures highlights the fact that while H3 is not the largest
spectroscopic survey, it is by far the largest medium-resolution
survey of distant high latitude stars, and rivals low-resolution
surveys of such populations.


\section{Summary}
\label{s:sum}

We have described a new stellar spectroscopic survey targeting the
stellar halo.  The goal is to trace the assembly history of the Galaxy
by studying the distribution of stars in 6D phase space plus
chemistry.  The survey aims to collect 200,000 stars sparsely sampled
over 15,000 sq. degrees.  The target selection is deliberately simple
and interpretable: a magnitude range of $15<r<18$ and {\it Gaia} DR2
parallaxes such that $\pi-2\sigma_{\pi}<0.5$ mas (recently updated to
$\pi<0.4$ mas).  As of June, 2019 we have collected 88,000 spectra.  The
\texttt{MINESweeper} stellar parameter pipeline is delivering RVs,
[Fe/H], [$\alpha$/Fe], and spectrophotometric distances for every
star.  A summary of the key parameters of the survey is provided in
Table \ref{t:h3}.  All of the data, including the derived stellar
parameters, will eventually be made publicly available via the survey
website: \texttt{h3survey.rc.fas.harvard.edu}.


\acknowledgments 

We thank the Hectochelle operators Chun Ly, ShiAnne Kattner, Perry
Berlind, and Mike Calkins, and the CfA and U. Arizona TACs for their
continued support of this long-term program.  Observations reported
here were obtained at the MMT Observatory, a joint facility of the
Smithsonian Institution and the University of Arizona.  This research
was supported in part by the National Science Foundation under Grant
No. NSF PHY-1748958.  CC acknowledges the hospitality of the KITP,
where this paper was written.  The computations in this paper were run
on the Odyssey cluster supported by the FAS Division of Science,
Research Computing Group at Harvard University.



\end{document}